\documentclass[a4paper,notitlepage,nofootinbib,amsmath,amssymb,prl,aps,longbibliography]{revtex4-1}
\usepackage{dcolumn}
\usepackage{bm}
\usepackage{hyperref}

\newcommand{\crt}{\\[2mm]}
\newcommand{\nn}{\nonumber}
\newcommand{\beq} {\begin{equation}}
\newcommand{\eeq} {\end{equation}}
\newcommand{\beqa} {\begin{eqnarray}}
\newcommand{\eeqa} {\end{eqnarray}}

\newcommand{\ie}{{\it i.e.}}

\newcommand{\as}{{\alpha_s}}
\newcommand{\lqcd}{\Lambda_{QCD}}
\newcommand{\la}{\Lambda}

\newcommand{\order}[1]{${\mathcal O}\left(#1 \right)$}

\newcommand{\eq}[1]{(\ref{#1})}

\newcommand{\gsim}{\agt}

\newcommand{\inv}[1]{\frac{1}{#1}}
\newcommand{\halft}{{\textstyle \frac{1}{2}}}

\newcommand{\quart}{{\textstyle \frac{1}{4}}}

\newcommand{\intt}{{\textstyle \int}}
\newcommand{\ket}[1]{\left\vert{#1}\right\rangle}
\newcommand{\bra}[1]{\langle{#1}\vert}

\newcommand{\tr}{\mathrm{Tr}\,}

\newcommand{\bs}[1]{\boldsymbol{#1}}

\newcommand{\mH}{\mathcal{H}}

\newcommand{\mS}{\mathcal{S}}

\newcommand{\xv}{{\bs{x}}} 
\newcommand{\yv}{{\bs{y}}}
\newcommand{\zv}{{\bs{z}}}

\newcommand{\Av}{{\bs{A}}}

\newcommand{\Ev}{{\bs{E}}}

\newcommand{\gz}{\gamma^0}

\newcommand{\nv}{\bs{\nabla}}

\newcommand{\rar}{\rightarrow}
\newcommand{\lar}{\leftarrow}

\newcommand{\rnab}{{\overset{\rar}{\nv}}\strut}
\newcommand{\lnab}{{\overset{\lar}{\nv}}\strut}

\newcommand{\alv}{{\bs{\alpha}}}

\begin{document}
\title{Bound states and perturbation theory\footnote{Talk at Light Cone 2019, 16-20 September 2019, Ecole Polytechnique, Palaiseau, France}}

\author{Paul Hoyer}
\email{paul.hoyer@helsinki.fi}
\affiliation{Department of Physics, POB 64, FIN-00014 University of Helsinki, Finland}
\homepage{http://www.helsinki.fi/~hoyer/}

\begin{abstract} 

A perturbative expansion for QED and QCD bound states is formulated in $A^0=0$ gauge. The constituents of each Fock state are bound by their instantaneous interaction. In QCD an \order{\alpha_s^0} confining potential arises from a homogeneous solution of Gauss' constraint. The potential is uniquely determined by the QCD action, up to a universal scale. The Cornell potential is reproduced for quarkonia, and corresponding ones found for higher Fock states, baryons and glueballs.

\end{abstract}


\maketitle

\parindent 0cm
\parskip.2cm

\section{Aspects of bound state dynamics}

Hadrons are bound states of QCD, with the binding energy scale set by $\lqcd \sim 200$ MeV. It is commonly thought that $\as(Q^2 \sim\Lambda_{QCD}^2) \gsim 1$, as would be the case if the perturbative running continued to low values of $Q$. However, the $Q^2$ dependence of $\as(Q^2)$ is established only for $Q \gsim m_\tau$, with $\as(m_\tau^2)\simeq 0.33$. $\as$ may be independent of $Q^2$ in soft dynamics, \ie, the coupling may freeze at a value which allows an expansion in powers of $\as(0)$. This would explain why hadron quantum numbers reflect their quark constituents, while glueballs and hybrids have yet to be found.  

Bound state dynamics differs qualitatively for strong and weak coupling. This is seen in QED$_2$, Quantum Electrodynamics in $D=1+1$ dimensions \cite{Coleman:1976uz}. The $e^-$ and $e^+$ form tightly bound bosons when their charge to mass ratio $e/m \gg 1$. The nearly pointlike neutral bosons interact weakly with each other and are the relevant degrees of freedom (d.o.f.) of the QED$_2$ dynamics at strong coupling. For weak coupling ($e/m \ll 1$) the QED$_2$ bound states are given by the Schr\"odinger equation, reflecting the fermion d.o.f.'s as for physical atoms. The strong coupling ($\alpha \gg 1$) spectrum of QED$_4$ is not known, but it appears unlikely that it would resemble atoms. 

In physical QED the Positronium wave functions are non-polynomial in $\alpha \simeq 1/137$, leading to non-perturbative features such as exponentially suppressed tunneling. The Positronium binding energies can nevertheless be perturbatively expanded in powers of $\alpha$ and $\log\alpha$, giving excellent agreement with data \cite{Adkins:2015wya}. The small value of $\alpha$ allows quantitative predictions even for strong external fields, as in the case of Schwinger pair production \cite{Schwinger:1951nm}.

Hadrons may be classified in terms of their quark constituents. The spectra and couplings of heavy quarkonia are well described by the Schr\"odinger equation with the Cornell potential  \cite{Eichten:1979ms,Eichten:2007qx},
\begin{align} \label{e1}
V(r) = V'r-\frac{4}{3}\frac{\as}{r} \ \ \ \text{with}\ \ V' \simeq 0.18\ \text{GeV}^2, \ \ \as \simeq 0.39
\end{align}
This potential was determined from fits to quarkonium data and later confirmed by lattice QCD \cite{Bali:2000gf}. The success of the Cornell approach indicates that the confinement scale $V'$ arises already at the classical (no-loop) level, with a gluon coupling $\as$ that is close to the perturbative $\as(m_\tau) \simeq 0.33$. Here we discuss a perturbative bound state expansion for hadrons, guided by a corresponding, first principles approach to QED atoms\footnote{A more detailed presentation may be found in \cite{Hoyer:2018hdj}.}. 

\section{Positronium from the QED action}

The Schr\"odinger equation is commonly derived by summing Feynman (ladder) diagrams, either explicitly or in terms of the Bethe-Salpeter equation \cite{Salpeter:1951sz}. The Feynman rules assume free ($in$ and $out$) states at asymptotic times, which excludes color confinement. We therefore consider a Hamiltonian approach. This can explain the origin of the linear term in the Cornell potential \eq{e1} for QCD. An analogous confining potential is not possible for QED.

We choose temporal ($A^0=0$) gauge \cite{Willemsen:1977fr,Bjorken:1979hv,Christ:1980ku,Leibbrandt:1987qv}, in which the absence of a conjugate field to $A^0$ is not an issue. The electric fields $E^i = F^{i0}=\partial_0 A_i$ are conjugate to the photon fields $A^i\ (i=1,2,3)$. The vanishing of Gauss' operator, defined by
\begin{align} \label{e2}
G(t,\xv) &\equiv \frac{\delta\mS_{QED}}{\delta{A^0(t,\xv)}} = \partial_i E^{i}(t,\xv)-e\psi^\dag\psi(t,\xv)
\end{align} 
is not an equation of motion in temporal gauge. $G(t,\xv)$ generates local gauge transformations which are time-independent and thus maintain $A^0=0$. The gauge is fully fixed by imposing Gauss' law as a constraint on physical states,
\begin{align} \label{e3}
G(t,\xv)\ket{phys} = 0
\end{align}
This determines the action of the longitudinal electric field $\partial_i E^{i} \equiv \partial_i E^{i}_L$ for each state.

The $e^+e^-$ Fock component of Positronium may be expressed as (henceforth $t=0$ is implicit),
\begin{align} \label{e4}
\ket{e^+e^-} = \sum_{\alpha, \beta}\int d\xv_1 d\xv_2\, \bar\psi_\alpha(\xv_1) \Phi_{\alpha\beta}(\xv_1-\xv_2) \psi_\beta(\xv_2)\ket{0}
\end{align}
where $\psi(\xv)$ is the electron field and the $c$-numbered wave function $\Phi(\xv_1-\xv_2)$ is a $4\times 4$ matrix in the Dirac indices $\alpha, \beta$. The state \eq{e4} has zero momentum since it is invariant under space translations. Only the electron and positron creation operators contribute ($b^\dag$ in $\bar\psi$ and $d^\dag$ in $\psi$).

Imposing the constraint \eq{e3} on the component $\ket{\xv_1,\xv_2}=\bar\psi_\alpha(\xv_1) \psi_\beta(\xv_2)\ket{0}$ gives,
\begin{align}
\partial_i E_L^{i}(\xv) \ket{\xv_1,\xv_2} &= e\psi^\dag\psi(\xv)\ket{\xv_1,\xv_2} = e\big[\delta(\xv-\xv_1)-\delta(\xv-\xv_2) \big] \ket{\xv_1,\xv_2}  \label{e5} \crt
E_L^{i}(\xv)\ket{\xv_1,\xv_2} &= -\frac{e}{4\pi}\,\partial_i^x\Big(\inv{|\xv-\xv_1|}-\inv{|\xv-\xv_2|}\Big)\ket{\xv_1,\xv_2} \label{e6}
\end{align}
The QED Hamiltonian in temporal gauge is
\begin{align} \label{e7}
\mH = \int d\xv\big[\halft(E_L^iE_L^i+E_T^iE_T^i)+\quart F^{ij}F^{ij}+\psi^\dag(-i\alv\cdot\nv-e\alv\cdot\Av+m\gz)\psi\big]
\end{align}
The longitudinal electric field $\Ev_L$ \eq{e6} contributes the potential energy of the $e^+e^-$ Fock state,
\begin{align} \label{e8}
\int d\xv\,\halft \Ev^2_L(\xv)\ket{\xv_1,\xv_2} = - \frac{\alpha}{|\xv_1-\xv_2|}\ket{\xv_1,\xv_2}
\end{align} 
The bound state condition $\mH\ket{e^+e^-}=(2m+E_b)\ket{e^+e^-}$ for the non-relativistic state \eq{e4} imposes the Schr\"odinger equation on the wave function $\Phi$. Fock states such as $\ket{e^+e^-\gamma}$, $\ket{e^+e^-e^+e^-}, \ldots$ contribute at higher orders in $\alpha$, and their instantaneous potentials can be determined similarly as for $\ket{e^+e^-}$. This perturbative Fock state expansion method in principle allows Positronium calculations of arbitrary accuracy. Explicit demonstrations are of course needed.

\section{Hadrons in QCD}

The $q\bar q$ Fock component of a meson state may be expressed similarly as for Positronium,
\begin{align} \label{e10}
\ket{q\bar q} = \inv{\sqrt{N_C}}\sum_{\alpha, \beta}\sum_{A,B}\int d\xv_1 d\xv_2\, \bar\psi_\alpha^A(\xv_1) \delta^{AB}\Phi_{\alpha\beta}(\xv_1-\xv_2) \psi_\beta^B(\xv_2)\ket{0} \equiv \int d\xv_1 d\xv_2\, \Phi(\xv_1-\xv_2)\ket{\xv_1,\xv_2}
\end{align}
The state is invariant under global gauge transformations since the wave function $\propto\delta^{AB}$ is a color singlet combination of the quark colors $A,B$. In the temporal ($A^0_a=0$) gauge of QCD \cite{Willemsen:1977fr,Bjorken:1979hv,Christ:1980ku,Leibbrandt:1987qv} the Gauss constraint \eq{e3} is
\begin{align} \label{e11}
\partial_i E_{L,a}^{i}(\xv)\ket{phys} = g\big[- f_{abc}A_b^i E_c^i+\psi^\dag T^a\psi(\xv)\big]\ket{phys}
\end{align}
For the $q\bar q$ Fock component in \eq{e10} we have at \order{g}, 
\begin{align} \label{e12}
\partial_i E_{L,a}^{i}(\xv)\ket{\xv_1,\xv_2} = g\bar\psi_A(\xv_1)T_{AB}^a\psi_B(\xv_2)\big[\delta(\xv-\xv_1)-\delta(\xv-\xv_2)\big]\ket{0}
\end{align}
In QED the component $\ket{\xv_1,\xv_2}$ of Positronium gives rise to the dipole electric field \eq{e6}. The color singlet meson state \eq{e10} cannot, however, generate an instantaneous color octet electric field $\Ev_{L,a}(\xv)$. The expectation value of $\partial_i E_{L,a}^{i}(\xv)$ in the color $C$ component of $\ket{\xv_1,\xv_2}$ is
\begin{align} \label{e13}
\bra{0}\psi_\beta^{C\dag}(\xv_2)\gz\psi_\alpha^C(\xv_1)|\partial_i E_{L,a}^{i}(\xv)|\bar\psi_\alpha^C(\xv_1)\psi_\beta^C(\xv_2)\ket{0} \propto g\big[\delta(\xv-\xv_1)-\delta(\xv-\xv_2)\big]T_{CC}^a
\end{align}
The color $C$ quark at $\xv_1$ feels the $\Ev_L$ field generated by its color $C$ antiquark partner at $\xv_2$, and {\it vice versa}. But an external observer does not experience a color field at \textit{any} $\xv$ since the sum over the quark colors $C$ vanishes, $\tr T^a = 0$. Hence we may consider solutions which (for each color component $C$) are non-vanishing at spatial infinity, without getting action-at-a-distance effects. We include a homogeneous ($\partial_i E_{L,a}^{i}(\xv)=0$) term in the solution of \eq{e11},
\begin{align} \label{e14}
E^i_{L,a}(\xv)\ket{phys} &= -\partial_i^x \int d\yv \Big[\kappa\,\xv\cdot\yv + \frac{g}{4\pi|\xv-\yv|}\Big]\mathcal{E}_a(\yv) \ket{phys}\crt
\mathcal{E}_a(\yv) &= - f_{abc}A_b^i E_c^i(\yv)+\psi^\dag T^a\psi(\yv) \nn
\end{align}
with a normalization $\kappa$ that is independent of $\xv$ and $\yv$. Since $\partial_i^x(\kappa\,\xv\cdot\yv)= \kappa\, y^i$ the field energy density of this (sourceless) term is independent of $\xv$, ensuring translation invariance. Together with rotational invariance this restricts the homogeneous solution to that given in \eq{e14}.

The QCD Hamiltonian in temporal gauge is
\begin{align} \label{e15}
\mH &= \int d\xv\big[\halft E_{L,a}^iE_{L,a}^i+\halft E_{T,a}^iE_{T,a}^i +\quart F_a^{ij}F_a^{ij}+\psi^\dag(-i\alv\cdot\nv+m\gz-g\alv\cdot \Av^a T^a)\psi\big]  
\end{align}
where $F_{ij}^a = \partial_i A_j^a-\partial_j A_i^a-gf_{abc}A_i^bA_j^c$\,. According to \eq{e14} the longitudinal electric field contributes
\begin{align} \label{e16}
\mH_V &\equiv \halft\int d\xv\,E_{a,L}^i E_{a,L}^i = \halft\int d\xv \Big\{\partial_i^x \int d\yv\Big[\kappa\, \xv\cdot\yv+\frac{g}{4\pi|\xv-\yv|}\Big]\mathcal{E}_a(\yv)\Big\}
\Big\{\partial_i^x \int d\zv\Big[\kappa\, \xv\cdot\zv+\frac{g}{4\pi|\xv-\zv|}\Big]\mathcal{E}_a(\zv)\Big\} \nn\crt
&= \int d\yv d\zv\Big\{\,\yv\cdot\zv \Big[\halft\kappa^2\intt d\xv + g\kappa\Big] + \halft \frac{\as}{|\yv-\zv|}\Big\}\mathcal{E}_a(\yv)\mathcal{E}_a(\zv) \equiv \mH_V^{(0)} + \mH_V^{(1)}
\end{align} 
where the terms of \order{g\kappa,g^2} were integrated by parts and $\mH_V^{(1)}$ denotes the \order{\as} gluon exchange contribution.

The components $\ket{\xv_1,\xv_2} = \sum_A \bar\psi_\alpha^A(\xv_1)\psi_\beta^A(\xv_2)\ket{0}$ are eigenstates of $\mH_V$,
\begin{align} \label{e17}
\sum_a\mathcal{E}_a(\yv)\mathcal{E}_a(\zv)\ket{\xv_1,\xv_2} = C_F \big[\delta(\yv-\xv_1)-\delta(\yv-\xv_2)\big] \big[\delta(\zv-\xv_1)-\delta(\zv-\xv_2)\big]\ket{\xv_1,\xv_2}
\end{align}
where $C_F = (N^2-1)/2N = 4/3$ for $N=N_C=3$. For $\mH_V^{(0)}$ in \eq{e16} this gives,
\begin{align} \label{e18}
\mH_V^{(0)}\ket{\xv_1,\xv_2} = C_F\big[\halft\kappa^2\intt d\xv + g\kappa\big] (\xv_1-\xv_2)^2\ket{\xv_1,\xv_2}
\end{align}
The \order{\kappa^2} contribution arises from the spatially constant field energy density, so it is proportional to the volume of space. It is irrelevant only if the energy density is identical for all bound state components. This determines $\kappa$ for the state $\ket{\xv_1,\xv_2}$ in terms of a universal constant $\la$,
\begin{align} \label{e19}
\kappa = \frac{\la^2}{gC_F}\inv{|\xv_1-\xv_2|}
\end{align}
The \order{g\kappa} and \order{\as} terms in \eq{e16} then give, respectively, the potentials
\begin{align}
V^{(0)}(|\xv_1-\xv_2|) &\equiv gC_F\kappa\, (\xv_1-\xv_2)^2 = \la^2 |\xv_1-\xv_2| \label{e20a} \crt
V^{(1)}(|\xv_1-\xv_2|)&=-C_F\frac{\as}{|\xv_1-\xv_2|} \label{e20b}
\end{align}
Neglecting higher Fock states the stationarity condition $\mH\ket{q\bar q} = M\ket{q\bar q}$ imposes a bound state equation on the wave function in \eq{e10},
\begin{align} \label{e9}
\big[i\alv\cdot\rnab+m\gz\big]\Phi(\xv)+ \Phi(\xv)\big[i\alv\cdot\lnab-m\gz\big] = \big[M-V(|\xv|)\big]\Phi(\xv)
\end{align} 
where $V = V^{(0)}+V^{(1)}$. In the non-relativistic limit this reduces to the Schr\"odinger equation and thus to the quarkonium model based on the Cornell potential \eq{e1} \cite{Eichten:1979ms,Eichten:2007qx}. The present approach is reminiscent of the Bag Model \cite{Chodos:1974je} in that the vacuum has a non-vanishing energy density. Yet there is no bag boundary, and the quarks move in the vacuum field which gives rise to the linear potential \eq{e20a}. 

At \order{\alpha_s^0} only the linear potential \eq{e20a} and the $q\bar q$ state \eq{e10} contribute, even for light quarks. The relativistic solutions of the bound state equation \eq{e9} are given in \cite{Hoyer:2018hdj}. Fock states with transverse gluons such as $\ket{q\bar qg}$ are generated by the Hamiltonian \eq{e15} at \order{g}. 

The instantaneous potential for any Fock state may be found using \eq{e14}. The field energy density, \ie, the \order{\kappa^2} term in the Hamiltonian $\mH_V$ \eq{e16}, must be the same for all states, making the scale $\la$ universal. Three examples  \cite{Hoyer:2018hdj}:
\begin{align}
\ket{gg} &= A_{a,T}^i(\xv_1)\,A_{a,T}^j(\xv_2)\ket{0}:\ V_{gg}(\xv_1,\xv_2) = \sqrt{\frac{N}{C_F}}\, \la^2\,|\xv_1-\xv_2|-N\,\frac{\as}{|\xv_1-\xv_2|}  \label{e21} \crt
\ket{qqq} &= \epsilon_{ABC} \psi_\alpha^{A\dag}(\xv_1)\,\psi_\beta^{B\dag}(\xv_2)\,\psi_\gamma^{C\dag}(\xv_1)\ket{0}: \ \text{With}\ \ d_{qqq}(\xv_1,\xv_2,\xv_3) \equiv \inv{\sqrt{2}}\sqrt{(\xv_1-\xv_2)^2+(\xv_2-\xv_3)^2+(\xv_3-\xv_1)^2}\ , \nn\\
&V_{qqq}(\xv_1,\xv_2,\xv_3) = \la^2 d_{qqq}(\xv_1,\xv_2,\xv_3)-\frac{2}{3}\,\as\Big(\inv{|\xv_1-\xv_2|}+\inv{|\xv_2-\xv_3|}+\inv{|\xv_3-\xv_1|}\Big)  \label{e22} \crt
\ket{qgq} &= \bar\psi_A(\xv_1)\,A_{b,T}^j(\xv_g)T^b_{AB}\psi_B(\xv_2)\ket{0}:\ \text{With}\ \ d_{qgq}(\xv_1,\xv_g,\xv_2) \equiv \sqrt{\quart(N-2/N)(\xv_1-\xv_2)^2+N(\xv_g-\halft\xv_1-\halft\xv_2)^2}\ , \nn\\
&V_{qgq} = \frac{\la^2}{\sqrt{C_F}}\, d_{qgq}(\xv_1,\xv_g,\xv_2)+\halft\,\as\Big[\inv{N}\,\inv{|\xv_1-\xv_2|}-N\Big(\inv{|\xv_1-\xv_g|}+\inv{|\xv_2-\xv_g|}\Big)\Big]  \label{e23}
\end{align}
In each case a bound state equation may be derived by adding the kinetic terms in the Hamiltonian, and the mixing with other Fock components taken into account at higher orders of $\as$.

\textit{Acknowledgements:}\\
I am grateful to the organizers of Light Cone 2019 for their invitation. During the preparation of this material I profited from a visit to  ECT* (Trento). I thank the Department of Physics at Helsinki University for my privileges of Professor Emeritus. A travel grant from the Magnus Ehrnrooth Foundation is highly appreciated.

\bibliography{LC2019_refs}
\end{document}